\documentclass[aps,pra,twocolumn,groupedaddress,amsmath,amssymb]{revtex4}

\usepackage{epsfig}
\usepackage{amssymb}
\usepackage{amsmath}
\usepackage{graphicx}
\usepackage{amsfonts}
\usepackage{epsfig}
\usepackage{revsymb}
\usepackage{bm}
\usepackage{amsmath}
\usepackage{amssymb}
\usepackage{latexsym}
\usepackage{thumbpdf}
\usepackage{hyperref}

\begin{document}

\title{On the transmission of crystallisation waves across the edge between the rough and faceted  crystalline surfaces in superfluid $^4$He }

\author{S. N. Burmistrov }

\affiliation{Kurchatov Institute, 123182 Moscow, Russia
\\  
}

\begin{abstract}
The wavelike processes of crystallisation and melting or crystallisation waves are well-known to exist at the crystal $^4$He surface in its rough state. Below the roughening transition temperature the crystal surface experiences the transition to the smooth faceted state and the crystallisation waves represent the propagation of a train of crystalline steps at the velocity depending on the crystal step height. Here we analyse the transmission and reflection of crystallisation waves propagating across the crystal edge separating the crystal surface in the rough and faceted states. 
\end{abstract}

\maketitle

\section{Introduction}

Helium crystals as a model system can provide us with very general and unusual properties of liquid-solid interfaces \cite{Bal05,Tsy15}. On one hand, helium crystals demonstrate faceting as classical crystals. The so-called roughening transition is the transition from the atomically rough and fluctuating state of the crystalline  surface at high temperatures to the smooth faceted  surfaces at sufficiently low temperatures. The experimental observations have displayed several roughening transitions in the hcp $^4$He crystals, as follows: $T_{R1}$=1.3 K for \textit{c}-facet in the [0001] direction, $T_{R2}$=1.07 K for \textit{a}-facet in the [10$\bar{\text{1}}$0] direction perpendicular to the \textit{c}-axis, and  $T_{R3}$=0.36 K for \textit{s}-facet in the [10$\bar{\text{1}}$1] direction. The [10$\bar{\text{1}}$1] direction is tilted by 58.5$^{\circ}$ with respect to the $[0001]$ direction. 
\par 
On the other hand, as compared with classical crystals, the $^4$He crystals under ultralow energy dissipation  can demonstrate the growth dynamics when quantum mechanics plays a major role \cite{Bal05,Tsy15}.  In particular, at sufficiently low temperatures the $^4$He crystal in contact with its superfluid phase can support oscillations of the superfluid-solid interface due to weakly damped processes of melting and crystallisation \cite{Kesh81}. From the dynamical point of view such weakly damped crystallisation waves at the rough crystal surface are an immediate counterpart of the familiar gravitational-capillary waves at the interface between two normal liquids and have the similar dispersion as a function of wave vector. 
\par
 On the contrary, no basic study of the melting-crystallisation dynamics has been made at the well-faceted and atomically smooth $^4$He crystal surfaces which, unlike the atomically rough crystal surfaces, have an infinitely large surface stiffness. Accordingly, the crystal surface curvature vanishes and the crystal facet takes the flat shape. The most striking distinction of smooth faceted crystal surfaces from the rough ones is  the existence of non-analytical cusplike behaviour in the angle dependence of the surface tension, e.g. Ref. \cite{Noz92}. The crystal step energy becomes nonzero and positive below the roughening transition temperature $T_R$, vanishing at the higher temperatures. The origin of the singularity is directly connected with nonzero magnitude of the facet step energy below the roughening transition temperature. 
\par 
As compared with the melting-crystallisation wave-like processes at the rough crystal surface, the analogous processes at the faceted crystal surface demonstrate a more complicated picture than those at the rough crystal surface \cite{Gus07}.  The frequency spectrum of crystallisation waves at the faceted crystal surface  has a sound-like dispersion with  the velocity depending significantly on the wave  perturbation amplitude and the number of facet steps distributed over the wavelength \cite{Gus07,Bur11}. In essence, such crystallisation waves represent a propagation of a train of crystal facet steps along the crystal surface at the velocity governed with the crystal step height.  Here we mention the formation of crystallisation waves under heavy shake of an experimental cell \cite{Kesh82} or in the process of anomalously fast growth of a $^4$He crystal under high overpressures \cite{Tsym00,Tsym04}.  The progressive facet waves are observed at the crystal (001) facet  in $^3$He \cite{Tse02}. 
\par 
The presence of singularity in the behaviour of surface tension or nonzero crystal step energy  results  also in a number of interesting phenomena at the faceted $^4$He crystal surface, e.g. amplitude-dependent velocity of traveling waves \cite{Gus07,Bur11}, quantum fingering of the inverted liquid-crystal interface in the field of gravity \cite{Bur08}, Rayleigh-Taylor instability with generating the crystallisation waves \cite{Bur09}, and electrohydrodynamical instability \cite{Bur12} with breaking the faceted state down.
\par 
So far the crystallisation waves have been studied only for the spatially homogeneous crystalline surfaces. 
Since the adjacent crystal surfaces have the different roughening transition temperatures, we can raise a question about the propagation of melting-crystallisation waves across the edge between the crystal surfaces in the rough and smooth states. For the first time, in the present paper  we attempt the transmission and reflection of crystallisation waves across the edge between the rough crystal surface  and the smooth faceted surface of a $^4$He crystal. 

\section{Lagrangian}
\par 
The atomically rough surface and the atomically smooth surface of a $^4$He crystal correspond to various crystallographic directions and the surfaces contact each other at the crystal edge. The transition from one direction to the other or from one surface to the other surface can be described with the polar angle which varies  gradually from one value to another in order to parametrise two adjacent crystal surfaces.
\par 
In order to treat  the transmission and reflection of crystallisation waves in most simplest and obvious way, we consider the following model situation. For simplicity, we assume that both the crystal surfaces, rough and smooth, are parallel to the $xy$ plane with the vertical position at $z=0$.  In addition, we imply that one-half of the crystal surface, e.g. $x<0$, is in the rough state and the other half $x>0$ is in the smooth faceted state. (Variable $x$ plays a role of polar angle.) First, we call $\zeta =\zeta (\bm{r})$ as a  displacement of the crystal surface from its horizontal position $z=0$ with $\bm{r}=(x,y)$ as a two-dimensional vector. We neglect any anisotropy of the crystal surface in the $xy$ plane as well. We suppose sufficiently low temperature range in order to neglect any possible  energy dissipation and  the damping of crystallisation waves at the both states of the surfaces. This implies the temperatures lower  than about 0.4~K. Neglecting the dissipation aspects simplifies mathematics as well. 
\par 
As a result, in the lack of energy dissipation the surface oscillations of a $^4$He crystal can be described with the following Lagrangian: 
\begin{eqnarray}
L[\zeta (t,\bm{r}),\dot{\zeta} (t,\bm{r})]=\frac{\rho _{\text{eff}}}{2}\iint d^2r\, d^2r'\frac{\dot{\zeta}(t,\bm{r})\dot{\zeta}(t,\bm{r}')}{2\pi |\bm{r}-\bm{r}'|} \nonumber 
\\ 
-\int d^2r\biggl(\alpha (\bm{\nu})\sqrt{1+(\nabla\zeta)^2}+\frac{1}{2}\varDelta\rho g\zeta ^2\biggr). 
\end{eqnarray}
Here we ignore the compressibility of the both liquid and solid phases and $g$ is the acceleration of gravity. Because of low-temperature consideration we shall also neglect the normal component density in the superfluid phase or, equivalently, the difference between the superfluid density $\rho _s$ and the density of the liquid phase $\rho$, i.e. $\rho _s=\rho$. Then the effective interface density $\rho _{\text{eff}}$ is given by 
$$
\rho_{\text{eff}}=\frac{(\rho '-\rho )^2}{\rho}\approx 1.9\,\text{mg/cm}^3
$$
and depends on the difference $\varDelta\rho =\rho '-\rho$ between the solid density $\rho '$ and the liquid density $\rho$. 
\par 
Unlike the fluid-fluid interface, the surface tension coefficient $\alpha (\bm{\nu})$ for the crystal depends essentially on the direction of the normal $\bm{\nu}$ to the interface against crystallographic axes. In our simplest description this is a function of the angle $\vartheta$ alone between the normal and, say, crystallographic [0001] or \textit{c} axis of the crystal hcp structure with the geometric relation $|\tan\vartheta |=|\nabla\zeta |$.  
\par 
For the crystal facet tilted by small angle $\vartheta$ from the basal plane, the expansion of the surface tension $\alpha (\vartheta)$, usually written (see Refs. \cite{Bal05,Noz92,Lan65}) as
$$
\alpha (\vartheta) =(\alpha _0+\alpha _1\tan |\vartheta |+\dots )\cos\vartheta ,\quad |\tan\vartheta |= 
|\nabla\zeta |, 
$$ 
can be expanded for the small angles into a series  
$$
\alpha (\vartheta)=\alpha _0+\alpha _1|\vartheta |+\ldots ,\quad |\vartheta |\ll 1. 
$$
We intentionally do not write the next terms of expansion, e.g., cubic ones due to step-step interaction,  since we are studying only a small bending of the crystal surface. The  rough or faceted state of the crystal surface is closely connected with the magnitude of $\alpha _1=\beta (T)/a$ representing a ratio of the linear facet step energy $\beta(T)$ to the crystallographic interplane spacing $a$. Below the roughening transition temperature $T_R$ the linear facet step energy $\beta (T)$ is positive and vanishes for temperatures $T>T_R$. Obviously, the dynamics of the rough and the faceted surfaces differs drastically in kind. 
\par 
To determine the spectrum of crystal surface oscillations, we minimise the action $S=\int L\, dt$ against perturbation $\zeta (t,\bm{r})$ in order to derive the equation of motion. Within the framework of our approximation $|\nabla\zeta |\ll 1$ the excess Lagrangian 
$\Delta L[\zeta , \dot{\zeta}]= L[\zeta ,\dot{\zeta}]-L[0,0]$ is given by the following expression: 
\begin{gather}
\Delta L[\zeta (t,\bm{r}),\dot{\zeta} (t,\bm{r})]=L[\zeta (t,\bm{r}),\dot{\zeta} (t,\bm{r})] -L[0,0]\nonumber 
\\ 
=\frac{\rho _{\text{eff}}}{2}\!\!\iint d^2r\, d^2r'\frac{\dot{\zeta}(t,\bm{r})\dot{\zeta}(t,\bm{r}')}{2\pi |\bm{r}-\bm{r}'|} \nonumber 
\\ 
-\!\int\! d^2r\biggl(\alpha _1|\nabla\zeta |+\frac{\alpha _0}{2}(\nabla\zeta)^2+\frac{1}{2}\varDelta\rho g\zeta ^2\biggr) . 
\end{gather} 
\par 
As for the step energy $\alpha _1$, we assume that its low temperature magnitude \cite{Bal05} is approximately $\alpha _1\approx 0.014$~erg/cm$^2$. This magnitude amounts to one-tenth of the surface tension $\alpha _0\approx$ 0.2 erg/cm$^2$ \cite{Bal05} and in the following we always keep inequality $\alpha _1/\alpha _0\ll 1$ in mind. Moreover, this small parameter justifies the approximations that will be made below.  
\par 
Note here that the faceted crystal plane represents in essence a region of the crystal surface in the rough-like state if it is tilted with respect to the crystallographic axis by the angle exceeding about $\arctan (\alpha _1/\alpha _0)\sim $ 4$^\circ$ in sense $\alpha _1|\nabla\zeta |\ll\alpha _0(\nabla\zeta )^2$. In fact, from the physical point of view the angle of slope $\vartheta=\arctan (\alpha _1/\alpha _0)\sim $ 4$^\circ$ is determined by the competition of two contributions into the total surface energy. One originates from the regular surface term $\alpha _0(\nabla\zeta )^2$ and the second does from the irregular step tension $\alpha _1|\nabla\zeta |$. Provided that $\alpha _0(\nabla\zeta )^2\gg \alpha _1|\nabla\zeta |$, the latter contribution becomes negligible and thus the dynamical interface properties should resemble those in the rough surface state.  One can say that the crystal surface has too many crystal steps. On the contrary, if $\alpha _0(\nabla\zeta )^2\ll \alpha _1|\nabla\zeta |$, the dominant linear term linear in $|\nabla\zeta |$ is responsible for faceting. 
\par 
Here we mean no phase transition from the atomically smooth to the rough state at about $\theta\sim 4^{\circ}$. We expect only that the dynamical response of the atomically smooth surface to its perturbation at sufficiently large tilted angles should resemble and becomes similar to the dynamical response of the surface in the rough state. The dependence of the dynamical response on the slope of the surface will represent the smooth crossover from one type of behaviour to another.  Such picture can be supported with the experimental evidence \cite{Rol94}. The crossover from the smooth to the rough-like state  is observed as a function of the tilt angle at the same magnitude between 3$^\circ$ and 4$^\circ$.  

\section{Transmission and reflection}
To consider the transmission and reflection of melting-crystallisation waves across the edge separating the rough surface and the crystal facet, we suppose a simple model to describe such phenomena. So, the step energy is approximated by the function 
\begin{eqnarray}\nonumber
\alpha _1(x)=\left\{
\begin{array}{cc}
0, & x<0
\\
\alpha _1 , & x>0.
\end{array}
\right.
\end{eqnarray}
 In other words, the left-hand side of the crystal surface is in the rough state and the right-hand side of the crystal surface represents the smooth faceted state. 
\par 
 The approximation for step energy $\alpha _1(x)$ with the step-like function implies implicitly that the width of transition $W$ from the rough to smooth faceted state is much smaller as compared with the inverse wave vector $1/k$ or wavelength. The smooth boundary when $W\sim 1/k$ or larger should change the transmission and reflection coefficients. The smooth transition usually reduces the reflection and enhances the transmission of the wave. 
\par  
 As the crystallisation wave propagates across the boundary between two crystal surfaces, the wave transmits and reflects.  The wave on the left-hand side of the boundary is a superposition of the incident and reflected waves. On the right-hand side from the boundary the transmitted wave alone propagates. The relation between all three waves is determined with the boundary conditions at the interface $x=0$.  We consider the case of the normal incidence. 
 \par 
Let us write the perturbations of the crystal surface due to the incident, reflected and transmitted waves, respectively,  
\begin{gather*}
\zeta _0(x,t)=\zeta _0e^{ikx-i\omega t},\quad x<0, 
\\
\zeta _1(x,t)=\zeta _1e^{-ikx-i\omega t}, \quad x<0, 
\\
\zeta _2(x,t)=\zeta _2e^{iqx-i\omega t}, \quad x>0.
\end{gather*} 
Here  $\omega$ and $k=k(\omega)$ are the frequency and wave vector of the incident and reflected waves. The transmitted wave has the same frequency but its wave vector $q=q(\omega ,|\zeta _2|)$, unlike the case of the rough state of the surface, depends on the amplitude of the wave as well. 
\par 
At the interface $x=0$ we should provide a continuity of both the crystal surface distortion $\zeta (x)$ and  the derivative of $\partial\zeta /\partial x$. Eventually, we choose the following boundary conditions in order to match the propagation of waves across the crystal edge: 
\begin{gather*}
\zeta (x=-0,t)=\zeta (x=+0,t),
\\
\frac{\partial\zeta (x=-0,t)}{\partial x}=\frac{\partial\zeta (x=+0,t)}{\partial x}\, .
\end{gather*}
These two natural boundary conditions allow us to hold for the finite magnitude of the total surface energy. 
\par
Employing these conditions, we obtain readily a pair of equations determining the reflection and transmission of the incident wave 
\begin{eqnarray*}
\left\{
\begin{array}{l}
\zeta _0+\zeta _1=\zeta _2, 
\\
ik\zeta _0-ik\zeta _1=iq\zeta _2. 
\end{array}
\right.
\end{eqnarray*} 
Introducing the reflection and transmission coefficients as a ratio of the reflected amplitude $\zeta _1$ and transmitted amplitude $\zeta _2$ to the incident one $\zeta _0$, we arrive at the following magnitudes: 
$$
r=\frac{\zeta _1}{\zeta _0}=\dfrac{k-q}{k+q}\quad\text{and}\quad t=\frac{\zeta _2}{\zeta _0}=\dfrac{2k}{k+q}. 
$$
\par 
The striking distinction from the usual case of acoustic wave is associated with the dependence of wave vector $q$ for the transmitted crystallisation wave on its amplitude $\zeta _2$.  The equations which determines the amplitudes $\zeta _2$ and $\zeta _1$ of the transmitted  and reflected waves are given by 
$$
\zeta _2=\dfrac{2k(\omega)}{k(\omega)+q(\omega ,\zeta _2)}\zeta _0, 
\quad\text{and}\quad \zeta _1=\zeta _2(\zeta _0)-\zeta _0 .
$$
\par 
To understand the main features of the phenomenon, we first neglect the gravitational term proportional to the density difference $\varDelta\rho$ in the Lagrangian, assuming that wave vector is larger than the inverse magnitude of capillary length $k>k_0\sim \sqrt{\varDelta\rho g/\alpha _0}$. Then, for the rough state of crystal surface, one has an ordinary capillary dispersion \cite{Kesh81}
$$
 \rho_{\text{eff}}\dfrac{\omega ^2}{k}=\alpha _0k^2\quad\text{and}\quad k(\omega)=\biggl( \dfrac{\omega ^2\rho_{\text{eff}}}{\alpha _0}\biggr)^{1/3}. 
$$
For the faceted state of the crystal surface \cite{Bur11}, the dispersion is more complicated and depends on the wave amplitude $\zeta$ according to 
\begin{eqnarray*}
\rho_{\text{eff}}\dfrac{\omega ^2}{q}=\left\{
\begin{array}{cc}
\gamma\alpha _1q/|\zeta |, & q|\zeta |\ll\alpha _1/\alpha _0,
\\
\alpha _0q^2, & q|\zeta |\gg\alpha _1/\alpha _0, 
\end{array}
\right.
\end{eqnarray*} 
where $\gamma=\pi\zeta (3)/7= 0.539\ldots$ is a numerical coefficient.  Accordingly, 
\begin{eqnarray*}
q(\omega ,\zeta)=\left\{
\begin{array}{cc}
\omega\biggl(\frac{\rho_{\text{eff}}|\zeta |}{\gamma\alpha _1}\biggr)^{1/2}\!\! , \; & \omega ^2|\zeta |^3\ll\frac{\alpha _1^3}{\alpha _0^2\rho _{\text{eff}}},
\\
\biggl(\frac{\omega ^2\rho_{\text{eff}}}{\alpha _0}\biggr)^{1/3}\!\! ,  \; &  \omega ^2|\zeta |^3\gg\frac{\alpha _1^3}{\alpha _0^2\rho _{\text{eff}}}.
\end{array}
\right.
\end{eqnarray*} 
\par 
The most interesting case is that of sufficiently small amplitudes $|\zeta _0|$ of the incident crystallisation wave satisfying the inequality $2k|\zeta _0|\ll (\alpha _1/\alpha _0)^{2/3}\lesssim 1$. The latter implies $q\ll k$. As a final result, we arrive at 
$$
\zeta _2\approx 2\zeta _0\, , \quad\zeta _1=\biggl(1-\frac{2q}{k}\biggl)\zeta_0\quad\text{and}\quad 
\frac{q}{k}\approx \sqrt{\frac{2\alpha _0}{\alpha _1}k|\zeta _0|}. 
$$
Thus, we have approximately the following reflection and transmission coefficients: $r\approx 1$ and $t\approx 2$. 
\par 
Let us discuss the result obtained.  We see that the reflected crystallisation wave has approximately the same amplitude and is similar to the incident wave but propagating in the opposite direction. At the same time the incident wave onto the boundary edge induces the transmitted crystallisation wave with the double amplitude representing the flat kink at the smooth crystal surface. Such soliton-like perturbation,  which size is about wavelength $2\pi /k$, propagates  away from the boundary at velocity of about $V\approx\bigl(\gamma\alpha _1/2\rho _{\text{eff}}|\zeta _0|\bigr)^{1/2}$. Briefly speaking, the incident wave produces the reflected wave and excites the transmitted wave in the shape of a soliton with the larger wavelength. 
\par 
In the opposite case of sufficiently large amplitudes, if $ 2k|\zeta _0|\gg (\alpha _1/\alpha _0)^{2/3}$, the reflected crystallisation wave is weak and practically vanishes since the facet step energy $\alpha _1$  plays a negligible role.  The reflection can appear only due to difference in the surface tension coefficients for the adjoint crystal facets: 
$$
r\approx\dfrac{\alpha _{0r}^{1/3}-\alpha _{0l}^{1/3}}{\alpha _{0r}^{1/3}+\alpha _{0l}^{1/3}}
$$
where coefficients $\alpha _{0l}$ and $\alpha _{0r}$ refer to the left- and right-hand sides of the crystal surface. In its turn, the transmitted crystallisation wave has an almost full similarity with the incident crystallisation wave. Thus, we expect $r\approx 0$ and $t\approx 1$ if  $\alpha _{0l}=\alpha _{0r}$.  

\section{Incidence from the crystal facet onto the rough crystal surface  }

\par 
Let us consider the opposite situation when the crystallisation wave or crystal step arrives at the boundary from the smooth faceted surface to the rough crystal surface. So, we represent the incident, reflected and transmitted waves as follows: 
\begin{gather*}
\zeta _0(x,t)=\zeta _0e^{-iqx-i\omega t},\quad x>0, 
\\
\zeta _1(x,t)=\zeta _1e^{iqx-i\omega t}, \quad x>0, 
\\
\zeta _2(x,t)=\zeta _2e^{-ikx-i\omega t}, \quad x<0.
\end{gather*} 
Then, we have the following conditions at the boundary $x=0$:
$$
\zeta _0+\zeta _1=\zeta _2\quad \text{and}\quad  -iq\zeta _0+iq\zeta _1=-ik\zeta _2\, .
$$
Hence we arrive at 
$$
\zeta _2=\dfrac{2q(\omega, \zeta _0)}{q(\omega ,\zeta _0)+k(\omega)}\zeta _0, 
\quad\text{and}\quad \zeta _1=\zeta _0-\zeta _2(\zeta _0).
$$
\par 
Again the most interesting case is when the amplitude of the incident wave is sufficiently small $q|\zeta _0|\lesssim\alpha _1/\alpha _0\ll 1$. This means, either the crystal step height is small, or the length of protrusive crystal layer is rather extended. So, we find that the wave vector of the transmitted wave is given by 
$$
k=q(\omega)\biggl(\dfrac{\gamma\alpha _1}{\alpha _0}\dfrac{1}{q(\omega)|\zeta _0|}\biggr)^{1/3}\gg q(\omega) 
$$
and the reflection and transmission coefficients read
$$
r=\frac{\zeta _1}{\zeta_0}\approx 1\quad\text{and}\quad t=\dfrac{\zeta _2}{\zeta _0}\approx2\biggl(\dfrac{\alpha _0}{\alpha _1}q\zeta _0\biggr)^{1/3}\ll 1. 
$$
Thus, we see that the crystallisation wave or the crystal step, on the whole, reflects from the rough crystal surface. As it concerns the transmitted wave, its amplitude is much smaller as compared with the amplitude of the incident wave and the excitation of the crystallisation wave at the rough surface with the incident crystal step is ineffective. Here we should underline some asymmetry between the incidence of crystallisation waves from the rough and from the faceted crystal surface sides. 
\par 
In the opposite limit when $q|\zeta _0|\gg \alpha _1/\alpha _0$ the difference between the faceted state and the rough state is not large. We expect practically no reflection from the boundary and the full transmission of the wave to the other side of the crystal surface. In fact, 
$$
k\approx q(\omega)+\dfrac{\gamma\alpha _1}{3\alpha _0|\zeta _0|}
$$
and the reflection and transmission coefficients are approximately given by 
$$
r=\dfrac{\zeta_1}{\zeta _0}\approx \dfrac{1}{6}\dfrac{\gamma\alpha _1}{\alpha _0q|\zeta _0|}\ll 1\quad\text{and}\quad t=\dfrac{\zeta_2}{\zeta_0} \approx 1. 
$$
The latter means that the noticeable reflection can again appear only due to large distinction in the surface tension coefficients of the left-hand and right-hand sides of a crystal. 

\section{Summary}
\par 
To conclude, for the first time we have attempted the transmission and reflection of crystallisation waves propagating in a $^4$He crystal across the boundary edge between the crystal surfaces in the rough and smooth states. The crystallisation wave at the rough $^4$He crystal surface  resembles the usual gravitational-capillary waves at the fluid-fluid interface. In contrast, the crystallisation wave at the smooth faceted surface in essence represents the propagation of crystal steps at the velocity depending on the crystal step height. To match two types of waves at the crystal edge, we use the natural boundary conditions. 
\par 
Since the dispersion of crystallisation waves at the smooth faceted surface is essentially governed with the wave amplitude, the transmission and reflection coefficients depend on the amplitude of the incident wave. 
The incidence of the crystallisation wave from the rough crystal surface onto the smooth faceted one  results in the practically mirror reflection of the incident wave and in inducing the crystal step or soliton   
of about double amplitude in the region of the crystal facet behind the crystal edge. 
\par 
In the opposite situation of the incidence of the crystallisation wave from the faceted crystal surface onto the rough crystal surface we should observe practically the full reflection and the corresponding small transmission to the rough crystal surface. Note that we have no symmetry with respect to rearrangement between the crystal surfaces in the rough and flat states. 
\par 
The experimental study on the dynamics of crystallisation waves at the atomically smooth crystal facet requires an effective mechanism for their excitation. Apparently this is a tough challenge. To confirm, we can mention an unsuccessful attempt to produce a soliton-like crystallisation wave with the aid  of a $\mathsf{\Pi}$-shaped crossbar oscillating in the vicinity of the crystal $^4$He facet \cite{Tsym13}. The oscillations of the crossbar are shown to be very effective for inducing the crystallisation waves at the rough $^4$He crystal surface but no effect is observed for the faceted $^4$He crystal surface. 
\par 
The present work proposes a mechanism for exciting the wave or soliton at the smooth flat facet with the help of crystallisation wave propagating along the non-faceted rough surface across the crystal edge in the direction to the atomically smooth facet. One more possibility is to prepare the electron-charged crystal facet in order to induce the instability of the atomically smooth surface at the critical electron density. However, this will require the larger critical electron density \cite{Bur12} by a factor of about 50 -- 100 as compared with that of about 10$^9$~cm$^{-2}$ observed for the rough interface. 
\par 
A special interest represents the experiment on the transmission and reflection of crystallisation waves propagating across the crystal edge between the rough crystal surface and the vicinal surface whose orientation is tilted by the small angle $\vartheta$ with respect to the well-faceted surface. Provided the tilt angle $\vartheta$ is sufficiently small, the crystal steps are well separated. On the other hand, an existence of ready-made crystal steps should noticeably affect the transmission and reflection coefficients as a  function of tilt angle of the vicinal surface. We hope an experimental production and examination of soliton-like crystallisation waves at the crystal $^4$He facet below the roughening transition temperature would be fascinating and incredible.  

\begin{acknowledgements}
The present study is performed with the pleasant collaboration with V.L. Tsymbalenko. The author is also grateful to L.B. Dubovskii. 
\end{acknowledgements}

\end{document}